\documentstyle[12pt]{article}
\newdimen\paperwidth
\newdimen\paperlength
\newdimen\margin
\newdimen\vmargin
\textwidth=\paperwidth
\advance \textwidth by -2\margin
\advance\hoffset by -1truein
\advance\hoffset by \margin
\textheight=\paperlength
\advance \textheight by -2\vmargin
\advance\voffset by -1truein
\advance\voffset by \vmargin
\advance\textheight by -2\baselineskip
\begin{document}

\renewcommand{\theequation}{\thesection.\arabic{equation}}
\newcommand{\Section}[1]{\section{#1}\setcounter{equation}{0}}
\newcommand{\SubSection}[1]{\subsection{#1}\setcounter{equation}{0}}

\begin{titlepage}
\title{ {\bf  Chaos in the Classical Analogue of the Hofstadter Problem}\thanks{
Work partly supported by CICYT under
contract AEN93-0776 (M.A. M.-D.) } }
%\thanks{e-mail: mardel@fis.ucm.es (m.a.m-d)}}

\vspace{2cm}   
\author{ {\bf M. del Mar Espinosa, Miguel A. Mart\'{\i}n-Delgado, Alfonso Niella,}\\
{\bf David P\'aramo and Javier Rodr\'{\i}guez-Laguna }\\
%\dag \mbox{$\:$} 
%and
%{\bf }\ddag \\
%\mbox{}    \\
{\em Departamento de F\'{\i}sica Te\'orica I}\\
{\em Universidad Complutense 28040-Madrid, Spain } }
\vspace{5cm}
\date{}
\maketitle
\def\baselinestretch{1.3}
\begin{abstract}

 The behaviour of an electron in a potential that resembles
that of a bidimensional solid with a perpendicular magnetic field applied
is studied from a classical point of view. This problem presents
the standard features of chaos and some new interesting patterns.
A new chaos indicator called {\em random walk indicator} is
presented to describe some of these new patterns.

\ \

\ \

\ \

\ \

\ \

\ \

\ \ 

\ \

\end{abstract}

\vspace{2cm}

\vskip-21.0cm
\rightline{UCM}
\rightline{{\bf July 1996}}
\vskip3in
\end{titlepage}

\newpage

   \Section{  Introduction}

Twenty years ago D.R. Hofstadter \cite{hofs} presented a celebrated work 
on the energy levels and wave functions of Bloch electrons on a two-dimensional
lattice in rational and irrational magnetic fields, hereafter refered to as 
{\em the Hofstadter problem}. The model involves a two-dimensional square lattice
of spacing $a$ immersed in a uniform magnetic field ${\bf B}$ perpendicular to it.
The electrons are treated in a tight-binding approximation which amounts to 
considering a discrete version of the two-dimensional Laplacian in the Schr\"{o}dinger
equation. The resulting eigenvalue equation becomes a finite-difference equation,
called Harper equation \cite{harper}, whose eigenvalues can be computed by a 
matrix method. The magnetic flux which passes through a lattice cell, divided by
a flux  quantum, yields a dimensionless parameter whose rationality or 
irrationality highly influences the nature of the computed spectrum. 
When the graph of the energy spectrum is plotted over a wide range of ``rational"
magnetic fields, a rich recursive structure is discovered in the graph. There are
large gaps whose form looks like a very striking pattern somewhat resembling 
a butterfly (the so called {\em Hofstadter butterfly}). Equally striking are the 
delicacy and beauty of its fine-grained structure. Likewise, the nature of the
energy spectrum at ``irrational" magnetic fields can also be deduced;
it is shown to be a Cantor set, i.e., an uncontable but measure-zero set of points.
Despite these features, it was also shown that the graph is continuous as the 
magnetic field varies.
All these fascinating features have led us to wonder what the possible behaviours
could be exhibited by a classical version of the Hofstadter problem as far as the
integrability and chaotic motion is concerned.

In this work we present a classical analogue of the Hofstadter problem along with
a detailed classical analysis of its integrability and onset of chaotic behaviour 
in several regimes. This is accomplished through the numerical study of four
standard chaos indicators: Lyapunov's exponent, 
 power spectrum,  Poincar\'e map
and correlation function.
Moreover, we present a novel chaos indicator which we call 
{\em random walk indicator} which we believe it is a suitable signal of chaos for
those systems exhibiting some kind of periodicity like the lattice periodicity of 
our model.

  The traditional way to tackle the study of a dynamic system
goes through the search of  constants of motion (first integrals) which
reduce the dimensionality of the problem. Liouville's Theorem
states that when we have  $N$  constants of motion in involution
(their Poisson brackets vanishing pairwise) the problem can be solved
through quadratures. When those $N$ constants do not exist the
system usually behaves erratically and suffers from instability in the initial conditions.
However, it is not easy to prove that those first integrals do not exist and in 
that case one has to resort to the study of some indicators which we have 
previously mentioned. It is admitted that there is clearly chaos when several of
these indicators show this to be the case \cite{berry}, \cite{berge}, \cite{ranada},
\cite{schuster}.

This paper is organized as follows. In Sect. 2  we introduce a possible classical
analogue of the Hofstadter model which is simple enough so as to allow a 
detailed analysis of its dynamics.
In Sect. 3 we present the results of our numerical study of the model based on
the computation of the standard indicators of chaos.
In Sect. 4 we describe the onset of a new pattern of chaotic motion which 
strongly resembles a random walk. This leads us to introduce a new chaos indicator
called {\em random walk indicator} to characterize this type of behaviour in 
systems with lattice periodicity.
In Sect. 5 we study the phenomenology of the model depending on the range of
the parameter space and the energy of the particle.
Section 6 is devoted to conclusions and prospectives.

    \Section{ Classical Analogue of the Hofstadter Model: Equations of Motion}

We shall be considering the classical 
motion of a charged particle, tipically an electron,
in a two-dimensional plane subject to the action of a static magnetic field of constant 
magnitud $B$
perpendicular to the plane. We also place this charged particle under the action
of a periodic potential given by,

\begin{equation}
V (x,y) = V_0 (\cos a x + \cos a y) \label{1}
\end{equation}

\noindent where $a$ is a parameter whose inverse plays the role of a 
{\em lattice spacing} in our model and $V_0$ sets the energy scale of the potential.
This potential looks like an {egg-crate} when pictured as in Fig. 1. This is the
way in which we model the presence of ions making up a two-dimensional crystal
at a classical level. We do not claim that this model is the precise 
classical limit of the Hofstadter model, but nevertheless our model captures the two
essential features exhibited by the quantum original model, namely, in-plane lattice
periodicity plus external magnetic field applied. It is in this sense that we call it
a classical analogue. Moreover, this type of periodic potential (\ref{1}) has also
been used in simple models of quantum solids, such as the Kronig-Penney model,
to mimic the presence of crystal ions. Thus, the model we propose to study is 
suitable and simple enough to study its classical behaviour.

The Lagrangian for our classical analogue is obtained by means of the minimal 
coupling with the electromagnetic gauge potential ${\bf A}$. The magnetic field
has components ${\bf B} = (0,0,B)$ and we choose the axial gauge to describe it, 
namely, ${\bf A} = {1\over 2}(-B x, B y, 0)$.
Thus, in Cartesian coordinates the Lagrangian reads as follows,

\begin{equation}
L = {1\over 2} (\dot{x}^2 + \dot{y}^2)- \tilde{V}_0 (\cos a x + \cos a y) + 
\tilde{B} (\dot{y} x - \dot{x} y) \label{2}
\end{equation}

\noindent where we have set the mass of the particle $m=1$ and introduced 
reduced quantities as $\tilde{V}_0\equiv q V_0$ and $\tilde{B}\equiv q B/2$,
$q$ being the charge of the particle.
The Euler-Lagrange equations of motion are,

\begin{equation}
\left\{ \begin{array}{l}
     \ddot{x} - B \dot{y} - V_0 \sin  x = 0 \\                                 \label{3}
     \ddot{y} + B \dot{x} - V_0 \sin  y = 0
\end{array} \right.
\end{equation}

\noindent where we have omitted the \~{}  to simplify the notation
henceforth and we have also set
$a=1$ so as to make $x=y=\pi$ a minimum. Equivalently, we could absorb the 
parameter $a$ into a redefinition of the $x,y$ coordinates and a further redefinition
of the potential strength $V_0$ by $V_0 a^2$. Thus, we are left with only two
independent external parameters, namely $B$ and $V_0$, characterizing our model.

\noindent The usual way to study the integrability of these equations of motion
(\ref{3}) is by searching for constants of motion according to Liouville's theorem.
To achieve this in a simple fashion one seeks the underlying symmetries of the
problem for then Noether's theorem guarantees us that an associated conserved 
quantity must exist. In our case the energy is conserved and it is given by,

\begin{equation}
     E = \frac{1}{2} (\dot{x}^2 + \dot{y}^2) + V_0 (\cos ax + \cos ay)           \label{4}
\end{equation}

\noindent 
We would need one more conserved quantity to render the model integrable, but as
it happens,
there are not further symmetries in our classical analogue model. For instance,
it lacks of central symmetry or rotational symmetry
 and therefore the angular momentum vector ${\bf J}$
is not a conserved quantity. There is
not even complete symmetry between x and y, for interchanging
them would reverse the sign of $\bf{B}$, which is an axial vector.
Therefore, the system has the first requisite so as to be a candidate to exhibit
chaotic behaviour in some regime of the parameter space.

Before getting into the numerical study of our model it is worth noticing that there
exist some approximation schemes wich lead to integrable regimes.
 These regimes are interesting in themselves as well as providing us 
with some guidance when we set off chaotic regimes in next section.

\noindent A first simple approximation is achieved if we set the electric potential
$V_0$ to zero, that is, we remove the solid picture of the model. In this case the
equation of motion simplify to,

\begin{equation}
\left\{ \begin{array}{l}
     \ddot{x} - B \dot{y} = 0 \\                                 \label{5}
     \ddot{y} + B \dot{x}  = 0
\end{array} \right.
\end{equation}

\noindent and the system has both translational and central symmetry. 
The trajectories are simply circles.

\noindent A second approximation is obtained by removing the magnetic field,
thus 

\begin{equation}
\left\{ \begin{array}{l}
     \ddot{x}  - V_0 \sin  x = 0 \\                                 \label{6}
     \ddot{y} - V_0 \sin  y = 0
\end{array} \right.
\end{equation}

\noindent we arrive at the equations for two uncoupled pendulums. The trajectories
are now deformed Lissajous curves (see Fig. 2).

\noindent Lastly, another approximation can be made while retaining both the electric
and  magnetic fields. It amounts to linearizing the original equation of motion 
(\ref{3}) as,

\begin{equation}
\left\{ \begin{array}{l}
     \ddot{x} - B \dot{y} - V_0  x = 0 \\                                 \label{7}
     \ddot{y} + B \dot{x} - V_0  y = 0
\end{array} \right.
\end{equation}

\noindent which turn out to be exactly solvable. The trajectories look like a 
flower (see Fig. 3) if the electric field is predominant $V_0 \gg B$ or like ``revolving
loop" (see Fig. 4) if the magnetic field dominates $B \gg V_0$.

The minimal coupling also allow us to write down the Hamiltonian of the system
in terms of the canonical momenta as 
$H={1\over 2} [(p_x+B y)^2+(p_y-B x)^2]+V_0 (\cos x + \cos y)$ 
and we could also perform a study of 
the associated Hamilton-Jacobi equation, but as the system is not integrable it
would not be very helpful either.

  \Section{ Standard Chaos Indicators: Numerical Calculations}

From the analysis of the previous section we have realized that the search for 
analytical solutions of (\ref{3}) is hopeless and the most convenient way to
proceed further in the study of our classical analogue model is by means of 
numerical calculations.

\noindent Prior to considering the computation of standard chaos indicators,
we have solved the equations of motion numerically for many different values of the 
external paramenters $V_0$ and $B$ and of the initial conditions to make sure
that our numerical methods performed well. We have employed a standard
numerical scheme such as the fourth-order Runge-Kutta coupled with a 
predictor-corrector method such as the Adams-Moulton method when it was 
necessary \cite{press}.
 We made sure that both the rounding  error and the natural error
(for numerical integration of Ordinary Differential Equations) were kept under
control by two different techniques: the systematic checking of energy conservation
and the checking of  ``inverse orbit integration", i.e., one must get the initial conditions
one departed from. In all these regards our model presents no problems.

Once we obtain the trajectory for a particular case of parameters and initial
conditions, we study the following chaos indicators which we briefly explain.
For an excellent introduction to standard chaos indicators the reader is refered to
\cite{ranadalibro} and also \cite{zugasti}, \cite{cuerno}, \cite{pgarcia} where it is
standard to use several indicators to test the presence of chaotic behaviour in a
classical system.
The detailed analysis of these data is left for the next sections.

     \SubSection{Poincar\'e Map}

This indicator constitutes a qualitative visualization of the topology of a 
classical dynamical system in the phase space. With its help it is possible
to visualize the torii associated to action-angle variables when they exist and 
their deformation and destruction due to non-integrable perturbations of the
Hamiltonian according to the precise description given by the KAM (Kolmogorov,
Arnold, Moser) theorem \cite{arnold}.

\noindent The Poincar\'e map is constructed as follows in our case. As our model
has two degrees of freedom its associated phase space is four-dimensional. 
Let us parametrize it with coordinates $(x,\dot{x},y,\dot{y})$. In fact, we do not 
need all that information for we also know that the motion takes place in the 
energy shell domain in this phase space due to energy conservation (\ref{4}).
Thus, we may get rid of one coordinate, say $y$. Next we may fix the value
of $\dot{y}$ (with e.g. $\dot{y}>0$) and hence study the phase motion in the 
plane (called Poincar\'e surface) 
defined by the remaining two coordinates $(x,\dot{x})$. 
The set of points resulting from the intersection of  a phase trajectory constitutes 
the Poincar\'e map. Recall that each phase trajectory is definided by giving a
value of the energy $E$ in (\ref{4}) as well as for the model parameters $V_0, B$.
For regular or integrable systems the Poincar\'e map is a well defined curve
for that means that another constant of motion would exist providing another 
constraint to eliminate the $\dot{y}$ coordinate in the same fashion as we did for $y$.
Example of this kind of regular behaviour in our classical analogue model can be 
found in Figs. 2, 5.
On the contrary, when the points of the Poincar\'e map fill some portion of the
Poincar\'e surface or are randomly scattered, this is an indication of chaotic
motion in the system (absence of enough first integrals to make the system
integrable). We have also found this kind of chaotic behaviour in our model
as can be seen in Figs. 6 a)-d) for different values of the external parameters and the
energy.

\SubSection{Lyapunov's Exponent}

One of the defining features of non-regular or chaotic motion is 
that the phase space gets  expanded  producing the separation of trajectories and
at the same time, the phase space folds into itself making the trajectories
to mix and blend among themselevs. The Lyapunov's exponent is a measure
of the first effect (expansion),
 or equivalently, the sensitive dependence on the initial conditions.
Namely, the Lyapunov's exponent $\lambda>0$ is computed preparing two trajectories
in the phase space which are very close initially. Then, we let them evolve a number
of $n$ time steps and compute again their separation. If their separation
grows exponentially as $e^{n \lambda}$, this is a possible indication of chaotic motion
provided the region of the phase space is bounded like is the case for 
energy-conserving systems. This condition is to guarantee that the expansion effect
is accompanied of the folding effect which altogether characterize the chaotic 
behaviour. For instance, a simple dynamical system such as $\dot{x}=x$ 
also exhibits exponential separation while being regular, but nevertheless it 
occurs in an unbounded region.

\noindent On the contrary, for regular bounded dynamical systems the separation
of nearby trajectories grows algebraically with time as $t^n$.

\noindent In Fig. 7 a) we plot our results for a regular behaviour while in 
Fig. 8 a) we show the case of chaotic motion. We plot the evolution with
time for the logarithm
of the separation of two initially very near trajectories divided
by the time interval. If it is high (positive) during an
appreciable lapse of time, the trajectories may have separated
exponentially.  This in turn would mean that  we shall have a high degree of
impredictability in our problem for we do not have infinite accuracy to determine
the initial conditions in practice.

   \SubSection{ Correlation Function}

     Another characteristic feature of a
 chaotic trajectory is the absence of any recognizable pattern in
its temporal evolution. 
Put in another way,
 it embodies more
information than a regular trajectory,  in the same way as the decimal
expansion of 1/7 has  less information than  $\pi$. 
The correlation function is a measure of whether a dynamical system 
retains the memory of its past history. To this purpose, we discretize one of 
the coordinates, say $x$, and define the correlation function $C_m$ for a 
$m$-step time interval as,

\begin{equation}
 C_m = {1\over n} \; \sum_{j=1}^n x'_j x'_{j+m}                         \label{8}
\end{equation}

\noindent where $x'_j\equiv x_j-\langle x \rangle=x'_j-\sum_k x_k/n$.
Thus, the correlation function indicates how similar is $x'$  (deviation of the coordinate
$x$ from its time average ) to its value $m$-time intervals after.
Therefore, a rapidly decreasing correlation funciton $C_m$ is an indication of chaos,
while if it remains finite it is and indication of regular motion.

\noindent In Fig. 7 b) we plot our results for a regular behaviour while in 
Fig. 8 b) we show the case of chaotic motion.

 \SubSection{ Power Spectrum}

The most useful and appropriate way to charectize a periodic (regular) motion is
through the analysis of its Fourier spectrum of frecuencies. Periodicity is translated
into the appearence of a peak at the characteristic frecuency of the dynamical
system. Quasiperiodicity amounts to a finite number of peaks corresponding to a 
set of principal frecuencies and their harmonics.
However, when we find a continuum spectrum we can suspect that the motion
occuring is very complex.

\noindent The power spectrum is defined as the intensity of the Fourier transform
for one discretized coordinate, say $x_j$. Namely, $E_k\equiv |\hat{x}_k|^2$.
    According to Wiener-Khinchin's theorem \cite{ranadalibro}, 
the correlation function and the 
power spectrum are related through a Fourier transform.

\noindent In Fig. 7 c) we plot our results for a regular behaviour while in 
Fig. 8 c) we show the case of chaotic motion.

 \Section{ A New Indicator: The Random Walk Indicator}

The analysis of the standard chaos indicators in the previous section shows us
that our classical analogue model is rich enough so as to exhibit both regular
and chaotic behaviour depending on the range of the parameter space $V_0, B$ 
and on the value of the energy of the system.
Furthermore, we shall show that our model also exhibits another type of behaviour
which we call random walk or brownian motion whose origin can be traced back
to the existence of a lattice periodicity in our system. This is the new ingredient
which allows us to introduce a new chaos indicator which we believe it is a
suitable one for systems presenting lattice periodicity of some sort.

Likewise the harmonic oscillator is the prototype of periodic system, a random
walk is one of the most aleatory dynamical systems when can think of.
A simple way to construct a random walk in the plane as shown in Fig. 9 a)
is by leaving a particle to hop in a two-dimensional lattice so that at each
time the particle chooses wheather to jump to the left-right or up-down by tossing
two coins. Remarkably enough, we have found a region in the parameter space
of our classical analogue model where the open trajectory underwent by the 
particle clearly looks like to that of a brownian motion as we show in Fig. 9 b).

\noindent Likewise the departure from the oscillator behaviour
is a sign of chaos which
is detected by the power spectrum, the proximity to a random walk can be considered
a sign of chaos as well. We have come up with a way to characterize this 
proximity by means of the random walk indicator. Thus, if the steps jumped
by a particle undergoing a random walk are of equal length, say one lattice spacing,
after $N$ steps the most probable position of the particle will be the initial
position. However, the mean square value of the position is just $N$. 
Let ${\bf r}(t)$ be the position vector of the particle in Fig. 9 b) at the time $t$ 
elapsed by hopping from the origin to a certain point of the total walk. 
Let us denote by $\langle {\bf r}^2\rangle$ the 
time average over the whole random walk in Fig. 9 b), namely,
$\langle {\bf r}^2 \rangle=\frac{1}{T}\int_0^T {\bf r}^2 dt$, $T\equiv N\Delta t$.
We may say that a random walk has a ``critical exponent'' $1/2$, because
r.m.s. distance grows as the square root of the elapsed time.
Therefore, we may introduce an indicator denoted by $I_{RW}$ as the quotient of 
the mean square value $\langle {\bf r}^2 \rangle$ to the number of total steps, i.e.,

\begin{equation}
I_{RW} \equiv {\langle {\bf r}^2 \rangle \over T}                       \label{9}
\end{equation}

\noindent 
Therefore, for a random walk $I_{RW}$ tends to a constant value equal to $1$
as $N\rightarrow \infty$.

\noindent 
To make the checking implicit in the random walk indicator it is required that the
the time elapsed be long enough so that the particle has left the unit cell of the
lattice and the space under observation  
must be also large enough so as to comprise several unit
cells for the hopping motion of the particle to make sense.
This is how the necessity for lattice periodicity  comes about.
Now,  when $I_{RW}$ tends to a constant value, as shown in Fig. 10 a) for a real
random walk, we may consider it as an indication of chaotic behaviour of this
kind. In this fashion, the plot of $I_{RW}$ for the trajectory depicted in Fig. 9 b) is
remarkably similar to that of the real random walk. To reinforce this point of view,
we have also plotted this indicator for the regular motion of Fig. 7 showing that
$I_{RW}\rightarrow 0$ as the characteristic feature of bounded motions in our 
scheme.

   \Section{  Analysis of Results: Phenomenology of the System}

In our search for non-usual behaviours in a classical system such as our classical
analogue model, we have found in the previous section the remarkable onset of
brownian motion when appropiate large time and space scales are considered for
a certain range of the paramter space $(V_0, B)$ and energy $E$. We have also
found standard regular motions and a variety of chaotic behaviour which we 
hereby describe in some more detail. We do this analysis depending on the value
of the energy chosen. Our favorite initial condition to start with has been launching
the charged particle from a minimum of the egg-crate potential (\ref{1}) towards
a maximum with a velocity given by the specified energy.

\SubSection{  Negative or Low Energies}

In this case the particle has not enough energy to scape and go to visit another
unit cell of the periodic potential (\ref{1}). Thus, the particle traces a sort of loops
inside one cell of a more or less complicated shape. We have found all sorts of 
them. From time to time, a ``peak"  in the trajectory (sudden change of direction) 
occurs. This happens when the particle approaches a saddle point of the potential.

     We have found chaos whenever  such "peaks" show up. It
comes out as a sort of ``line broadening" of the curve forming the Poincar\'e map.
This is the first step towards the filling of a whole region signalling chaos.
 For a given $B$ there
is a minimum value for $V_0$ below which there is no chaos, and
there is also an upper value for $V_0$ above which there is
regularity  ever after.
 There are slow-motion regular trajectories which are fine, but there also exist
rapid-oscillatory motions which we call ``soft chaos", for they
present high Lyapunov exponent, but a correct (regular) Poincar\'e map  and
 power spectrum.

We have collected all our qualitative data in the schematic Table 1 
showing the qualitative behaviours found for $E=0$ 
(Low Energy regime) and several values of the external parameters $(V_0, B)$.
     Maps for lower energies are alike, but there is no chaos and there is
only the previously mentioned broadening of lines in the Poincar\'e maps. 
     Another interesting feature is the appareance of a``non-uniform space filling"
of the Poincar\'e surface as depicted is Figs. 6 a) and b). Thus,
chaos is present as usual when a full area is filled.

\SubSection{  High Energies}

     For these values of the energy the trajectory is bounded if the
magnetic field is  high enough, that is to say,  when the intensity of
the cyclotron frequency prevails in the Fourier spectrum. As a
consequence, the motion is  close to regular.

     Another feature for this energy regime is that the trajectory
has always a sensitive dependence on the initial conditions for  the
electron crosses several cells of the periodic potential, 
so it can pass close to unstable
points, such as the maxima of the egg-crate potential (\ref{1}) (if
energies were lower, those points would not be reached). The Lyapunov
exponent is then high and steadily growing. 
     In this case the Poincar\'e map as it has been defined in
Sect. 3 a) is a bad illustration because
it is difficult to discern between line-broadening case and a space-filling case
due to the fact that the particle travels many unit cells. Thus, we have chosen to
work in a reduced Poincar\'e surface consisting of just one unit cell due to the 
periodicity of our model (i.e., we apply periodic boundary conditions converting the
space into a torus). This makes the problem of drawing an easier one for now the
Poincar\'e map is bounded. Otherwise it covers an indefinite region in the phase 
space and it is difficult to classify the different patterns of behaviour.
In summary, we have found the following patterns:

\begin{itemize}

\item  {\em Chaotic motion:} it appears very early when the particle has 
enough energy to visit several cells. Regular trajectories are
associated with a very low or very high value for any of the
parameters $V_0$ or $B$ (there must
be hegemony of one parameter upon the other in order to have regularity).  
 
\item  {\em Temporally bounded trajectories:} a particle orbits around
a minimum of the potential  for say 18 loops and when one thinks 
it has been captured,
it scapes and wanders erratically around
the lattice. Eventually, it behaves like a random walk as in Fig. 9 b). 

\item {\em Mixed Complex Behaviour:}
Here almost all chaos signals ``light on" simultaneously: Poncar\'e
maps with space filling, high Lyapunov exponents and random
walks. Nevertheless, the correlation function keeps high values for the particle 
behaves in 
almost the same fashion in every cell it visits, with only some peaks,
loops or sudden changes in direction from time to time.
     
\end{itemize}

\noindent Likewise, we illustrate
 all our qualitative data in the schematic Table 2 
showing the qualitative behaviours found for $E=12$ 
(High Energy regime) and several values of the external parameters $(V_0, B)$.

       \Section{Conclusions}

We have presented in this paper a possible classical analogue of the Hofstadter
model \cite{hofs} described in the introduction along with a detailed study of a 
great variety of classical dynamical behaviours exhibited by our simple model.
Some of the patterns found for the Poincar\'e maps described in the last section
we believe are specific of a system with spatial lattice periodicity such as ours.
We have also employed other standard chaos indicators to carry out this analysis.
Moreover, the finding, in a certain range of the parameter space $(V_0, B)$
of a pattern of behaviour which strongly resembles a random
walk or brownian motion, see Fig. 9 b), has led us to introduce an additional
chaos indicator to check the onset of a random walk behaviour in systems 
with some sort of lattice spacing.

In this present work we have been clearly working at a very descriptive level
and we believe that a more thorough understanding of the intricacies of these
types of models is necesary. For instance, the connection between the classical
and quantum versions of the Hofstadter model deserves more attention regarding
the relation between chaos and quantization of Hamiltoian systems which is an
open research field \cite{ozorio}.

All these considerations make us believe that the study of classical dynamical 
systems with some sort of lattice periodicity might quite well embody new features
not present in standard studies of chaotic behaviour.

\vspace{30 pt}

     {\bf Acknowledgements}

We would like to thank  A.F. Ra\~{n}ada for his encouragement
to carry out this work and useful comments.
We also wish to thank E. Olmedilla for providing us with  access to the 
computer facility {\em Aula de Informatica} in the Departamento de 
F\'{\i}sica Te\'orica, Universidad Complutense.

\noindent
Work partly supported by CICYT under
contract AEN93-0776 (M.A. M.-D.). 

\noindent
This work was done while four of the authors (M.E.E., A.N., D.P. and J.R.-L.)
where attending an undergraduate course in theoretical mechanics imparted
by the remaining author (M.A.M.-D.) and A.F. Ra\~nada in the Universidad
Complutense de Madrid.

E-mail: 
javirl@sisifo.imaff.csic.es,
mardel@eucmax.sim.ucm.es

\newpage
\section*{Table captions}

 {\bf Table 1 :} Schematic table showing the qualitative behaviours found for $E=0$ 
(Low Energy regime) and several values of the external parameters $(V_0, B)$.
The meaning of the symbols are: 
 $\bullet =$ Space-filling in the Poincar\'e map,
    $\Box =$ Non-Uniform space filling,
  $\| =$ Line-Broadening.
    P = Quasiperiodic motion.

{\bf Table 2 :} Schematic table showing the qualitative behaviours found for $E=12$ 
(High Energy regime) and several values of the external parameters $(V_0, B)$
(the meaning of the symbols are as in table 1).

\newpage
\section*{Figure captions}
\noindent

 {\bf Figure 1 :} Picture of the periodic potential in Eq. (\ref{1})
 employed to mimick a crystal solid
in our classical analogue of the Hofstadter model. 

 {\bf Figure 2 :}  A typical trajectory in configuration space for $B=0$ along with
its Poincar\'e map.

 {\bf Figure 3 :} Flower-like trajectory for the linear approximation when $V_0 \gg B$.

 {\bf Figure 4 :} Revolving-loop trajectory for the linear approximation when $B \gg V_0$.

 {\bf Figure 5 :} Poincar\'e map for $V_0 = 3, B = 1, E = 0$.

 {\bf Figure 6 :} a) Poincar\'e map for $V_0 = 16, B = 5, E = 2$.
                       b) Poincar\'e map for $V_0 = 7, B = 4, E = 0$.
                       c) Poincar\'e map for $V_0 = 4, B = 3, E = 12$.
                       d) Poincar\'e map for $V_0 = 4, B = 6, E = 12$.

 {\bf Figure 7 :} Plots of some standard chaos indicators for $V_0 = 3, B = 1, E = 0$.
                     a)  Lyapunov exponent. 
                      b) Correlation function.
                      c) Power spectrum.

 {\bf Figure 8 :} Plots of some standard chaos indicators for $V_0 = 7, B = 4, E = 0$.
                     a)  Lyapunov exponent. 
                      b) Correlation function.
                      c) Power spectrum.

 {\bf Figure 9 :}  a) Picture of a real two-dimensional random walk.
                        b) Plot of the random walk appearing in our classical analogue 
                           model for $V_0 = 16, B = 5, E = 2$.

 {\bf Figure 10 :} Plot of the random walk indicator $I_{RW}$ for:
                    a) a real two-dimensional random walk.
                    b) the random walk appearing in our classical analogue 
                           model for $V_0 = 16, B = 5, E = 2$.
                     c) a regular motion appearing for $V_0 = 3, B = 1, E = 0$.

\newpage

\begin{table}[p]
\centering
\begin{tabular}{|c||c|c|c|c|c|c|c|c|c|c|c|c|c|c|c|}  \hline \hline
$B \ V_0$ &  1 & 2&  2.05&  2.5&  3&  4&  4.5&  5&  6&  7&  8&  9&  10& 11& 12 \\ 
\hline   \hline
2    &          P &  P &    P &     P&   P & P&   P&   P&  P&  P&  P&  P&  P&  P&  P \\
\hline 
2.5 &P& P& $\|$ & $\bullet$ & $\bullet$ & P&   P& P&  P&  P&  P&  P&  P& P& P \\
\hline
3& P&P& P&P& $\bullet$ & $\bullet$ & $\| $ &P&P& P&P&P&P  & P&  P \\ \hline
4&P&P&P&P &P&P &$\bullet$ & $\bullet$ &  $\bullet$&$\Box$&$\|$&  P&P&P&P \\
\hline
5&P&P&P&P&P&P&P&P&P&P& $\|$& $\bullet$ & $\bullet$& $\bullet$ & $\bullet$ \\
\hline            
6&P&P&P&P&P&P&P&P&P&P&P&P&P&  $\|$&    $\bullet$   \\ \hline \hline
\end{tabular}
\caption{Schematic table showing the qualitative behaviours found for $E=0$ 
(Low Energy regime) and several values of the external parameters $(V_0, B)$.
The meaning of the symbols are: 
 $\bullet =$ Space-filling in the Poincar\'e map,
    $\Box =$ Non-Uniform space filling,
  $\| =$ Line-Broadening.
    P = Quasiperiodic motion.} 
\end{table}

\begin{table}[p]
\centering
\begin{tabular}{|c||c|c|c|c|c|c|c|c|c|c|c|}  \hline \hline
$B \ V_0$ &  1 & 2&  4&  5&  6&  7&  8&  9&  10& 11& 12 \\ 
\hline   \hline
0.1 &  $\Box$ &  P& P& P& P&P& P& P& P& P&  P \\ \hline  
0.5 &$\bullet$ & $\bullet$ &  $\bullet$&$\bullet$ & $\bullet$ &  $\bullet$
&$\bullet$ & $\bullet$ &  $\bullet$&$\bullet$ & $\bullet$ \\ \hline                                                    
1&$\|$&$\|$ &$\bullet$ & $\bullet$ &  $\bullet$&$\bullet$&P
&$\bullet$ & $\bullet$ &  $\bullet$&$\bullet$ \\ \hline                    
2&P$\|$&P$\|$&$\bullet$ & $\bullet$ &  $\bullet$&$\bullet$ &$\bullet$
&P &   P &   P &   P \\ \hline
3&P$\|$&P&P&$\bullet$&P&P&P&P&P&P&P  \\ \hline
4&P&P&P&P&    P &   P&    P&    P&    P&    P&    P  \\ \hline
5 &P&P&P&P&    P &   P&    P&    P&    P&    P&    P  \\ \hline \hline
\end{tabular}
\caption{Schematic table showing the qualitative behaviours found for $E=12$ 
(High Energy regime) and several values of the external parameters $(V_0, B)$
(the meaning of the symbols are as in table 1).} 
\end{table}

\end{document}